\documentclass{cimento}

\usepackage{graphicx}
\usepackage[font=footnotesize, format=hang, labelfont={bf}]{caption}
\usepackage{float}
\usepackage{booktabs}
\usepackage{caption}
\usepackage{subcaption}

\usepackage{amsmath}
\usepackage{amssymb}
\usepackage[version=3]{mhchem}
\usepackage{bm}
\usepackage{nicefrac}

\newcommand{\bb}{$0\nu 2 \beta$} 

\newcommand{\mbb}{m_{\beta \beta}} 

\newcommand{\taubb}{\mbox{\normalsize $T_{\mbox{\tiny $\nicefrac{1}{2}$}}$}} 

				\newcommand{\GeV}{\text{GeV}}
\newcommand{\TeV}{\text{TeV}}

		\newcommand{\cm}{\text{cm}}	

\newcommand{\yr}{\text{yr}}	\newcommand{\kg}{\text{kg}}	\newcommand{\sss}{\text{s}}	\newcommand{\sr}{\text{sr}}

\newcommand{\el}{\text{e}}	\newcommand{\pr}{\text{p}}

\title{La Thuile 2014: Theoretical premises to neutrino round table}
\author{Francesco Vissani
	}
\instlist{\inst{}
	INFN, Laboratori Nazionali del Gran Sasso 
	{\em and} Gran Sasso Science Institute, L'Aquila, Italy}

\begin{document}

	\maketitle

	\begin{abstract}
		This talk, dedicated to the memory of G.~Giacomelli, introduced the round table on neutrinos held in February 2014. The topics selected for the discussion are: 1)~the neutrinoless double beta decay rate
		(interpretation in terms of light neutrinos, nuclear uncertainties); 2)~the physics in the gigantic water Cherenkov detectors (proton decay, atmospheric neutrinos); 3)~the study of neutrino oscillations (mass hierarchy and CP violation; other neutrino states); 4)~the neutrino 
		astronomy at low and high energies (solar, supernova, cosmic neutrinos).    
		The importance of an active interplay between theory and experiment is highlighted.
	\end{abstract}


	Neutrino physics is in a healthy state.
	Among the reasons of this pleasant situation, there are  the important results that have been achieved in 
	that field, as well as the possibility of conceiving/conducting new valuable experiments, the numerous
	installations and experimental sites, the 
	available sources of support and funding, etc.\,. 
	One specific circumstance that  contributes to make neutrino physics still valid and appealing is the vibrant connection between theory and experiment. 
	This is something that derives us from 
	the early times of neutrino physics and that 
	gives rise to a lively, sometimes messy, interdisciplinary field, where nuclear physics, 
	particle physics and astrophysics are involved. 
	In order to stress the importance of this connection in the way that it deserves, 
	I would like to dedicate this talk to Giorgio Giacomelli, one of the best experimentalists 
	we ever had, always open to discuss, to assess the value, and whenever necessary, to test the best ideas.
	Moreover, I open the list of references
	just as in my previous talk at La Thuile, namely,  
	with the review article of B.~Pontercorvo \cite{bruno}  that summarizes  the history of neutrino physics before 1983 
	in 4 tables, whose titles are,
	\begin{itemize}
		\item[I:]	from radioactivity to neutrino discovery;
		\item[II:]	from muon properties to V-A;
		\item[III:]	from high-energy neutrinos \footnote{This is the same of ``artificial (or long baseline) neutrino 
			beams'' in the modern parlance.} to the standard model;
		\item[VI:]	neutrino astrophysics, astronomy, cosmology.
	\end{itemize}
	Most of the lines of research listed in these four tables   
	are still pursued, and among them we note:
	\begin{itemize}
		\item[I:]	\emph{Neutrinoless double beta decay}
		\item[II:]	\emph{Neutrino oscillations}
		\item[III:]	\emph{Proton decay?} (the question mark is as in Pontecorvo's review.)
		\item[VI:]	\emph{Solar, supernova, high-energy cosmic neutrinos}.
	\end{itemize}
	In the following, 
	we will recall the main facts occurred since Pontecorvo's review article  and mention some possibilities of further development.

\section{Neutrinoless double beta decay}

	In this section, we  describe the present theoretical understanding of  a hypothetical nuclear physics process, namely, 
	 the neutrinoless double beta decay (\bb) process, 
	\begin{equation}
		(A,Z) \to (A,Z+2) + 2\el^-
	\end{equation}
and consider the most relevant pending questions.

	The existing experimental results have been summarized by M.\ Agostini, with particular emphasis on the investigations 
	using \ce{^{76}Ge}. The study of this nuclear species allows us to exclude the occurrence of \bb~transition with 
	half-life smaller than few times $10^{25}\,\yr$ after collecting a total exposure nearing $100\,\kg\cdot \yr$ for 
	\ce{^{76}Ge}. \footnote{This limit was obtained by combining the entire set of experiments. We remind that: 1)~a subset 
	of the 
	Heidelberg-Moscow collaboration claimed the discovery, but the claim is not confirmed; 
	2)~for \ce{^{136}Xe} exposure and background are both larger; the results are similar to \ce{^{76}Ge}.} 
	Although this exposure is about one million times smaller than those collected in proton decay search, it is quite remarkable since it concerns a rather peculiar nuclear species.
	
\subsection{Importance of the process}

From a phenomenological point of view, we can describe  \bb~as a nuclear decay  in which a pair of leptons (electrons) are generated.
In other words, the \bb\  is an example of lepto-genesis process potentially measurable in the laboratory. 

	The \bb\ is forbidden in the Standard Model (SM) of elementary particles by the conservation of the lepton number $L$.
	But $L$ is just an accidental symmetry of the SM. In the more complete models that respect the gauge symmetry of the SM, and that allow for non-zero neutrino masses, it is typically violated. 
	
In the SM, the lepton number $L$ and the baryon number $B$ (or equivalently the quark number $Q=B/3$) have the same theoretical status. 
Therefore, the search of \bb\  has a comparable importance to the search of proton decay, another hypothetical process where, this time, the baryon number $B$ 
is instead violated (see \textsection~\ref{sec:pdecay}). 
(Recall that the processes of creation  and of destruction are 
deeply linked in quantum field theory.)

	The matter is made of atoms, or at a more fundamental level,  
	of baryons (quarks) and leptons. These fermions are generically referred to as `matter', in contrast with the bosons that are the particles that carry the  forces. In this sense, we can say  the hypothetical 
	\bb\ and the proton decay are processes where the {\em matter} is created and 
	destroyed (while, of course, {\em mass} is neither created in the \bb\ process nor destroyed in the proton decay).	 
	 The observation of \bb~or of proton decay, or, even better, of both, would allow us to proceed toward a dynamical explanation of the cosmic baryon excess, a theoretical program first formulated by A.\ Sakharov \cite{sak}.

	There are   
	non-perturbative quantum processes that respect $B-L$ but 
	violate the symmetry $B+L$ \emph{within} the SM~\cite{krs}. 
	Since \bb\ violates $B-L$,
	one concludes that 
	1)~this process is forbidden in the SM also at 
	a non-pertubative level;
	2)~its observation would imply the existence of transitions 
	that violate also $B$ at some level \cite{krs}.   
	A link with the cosmological considerations mentioned above is quite plausible; indeed, it exists in several specific theoretical realizations of the program of Sakharov.

	\begin{figure}[t]
		\centering
		\includegraphics[width=8.7cm]{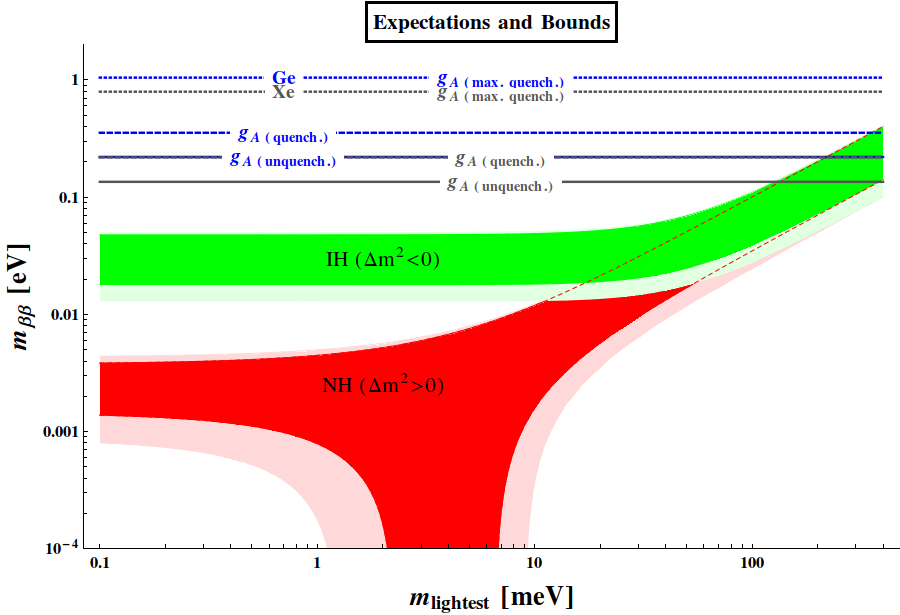}
		\caption{Expectations on the parameter $m_{\beta\beta}$
		from three flavor oscillations  \cite{ssv}. 
			In the red region we assume normal mass hierarchy; in the green one inverted mass hierarchy. The error bands 
			($3 \sigma$) are due to the incomplete knowledge on the oscillation parameters. The  
			experimental bounds (horizontal lines) depend on the assumption on the value of $g_A$.}
		\label{fig1}
	\end{figure}

\subsection{Connections with oscillations of ordinary neutrinos\label{merlos}}

	The discovery of neutrino oscillations (see \cite{lt2003,rev} for reviews and  \textsection~\ref{nooco} for further discussion)
	and the consequent inference that the three ordinary neutrinos are endowed 
	with mass, allow us to discuss in more concrete terms the rate of this process and to test the 
	theoretical proposal of E.\ Majorana, namely, that neutrinos are real particles, just as the photon. 
	More precisely, we wonder whether the neutrino mass eigenstates are real particles, which means to ask whether 
	the masses $m_i\ge 0$ are included in the Lagrangian as $\sum_i m_i \nu_i^t C^{-1} \nu_i/2 + h.\,c.$\,.
	The combination of the three neutrino masses entailed by the \bb~decay process is 
	\begin{equation}
		\mbb =\left|  \sum_{i=1}^3 U_{ei}^2\ m_i \right|
	\end{equation}
	where $U_{ei}$ are the mixing elements that define the electron neutrino state: 
	$|\nu_e \rangle =U_{ei}^* |\nu_i \rangle$. Unfortunately, oscillations probe only the absolute values of 
	$|U_{ei}|$, the squared mass difference between the closest mass levels 
	$m_2^2-m_1^2$; the remaining mass difference $m_3^2-m_1^2$ is known up to its sign.  
	In other words, we do not 
	known whether $m_3$ is the heaviest state (normal hierarchy) or the lightest one (inverted hierarchy). 
	Thus, a wide range of values for $\mbb$ are allowed, as shown in Fig.\ \ref{fig1} and 
	as discussed in details in  \cite{ssv}.
 	
	A priori, it is impossible to exclude the presence of \emph{other} contributions to the transition rate. 
	If the transition was observed, but the connection indicated by Fig.~\ref{fig1} was not viable, an  
	alternative interpretation would be necessary.
	But the first non-renormalizable operator that violates the lepton number implies Majorana neutrino masses and 
	nothing else~\cite{wei}.
	Thus, if the physics that explains neutrino masses is at very high energies and the subsequent (higher dimensional) operators are very suppressed, a one-to-one connection between the transition rate of the \bb~and the Majorana masses holds true. 

\subsection{Nuclear physics aspects}

	A major difficulty in the interpretation of this results arises when we convert the limit on the lifetime in terms of 
	limit on the parameter $\mbb$, since this requires to describe precisely the nuclear structure of initial, final and 
	intermediate nuclear species. This is summarized into a single adimensional parameter, the nuclear matrix element of 
	the transition $\mathcal{M}$. The half-life of the transition (connected to the rate 
	$\Gamma$ by $\taubb=\ln(2)\, \hbar/\Gamma$) is given by:
	\begin{equation}
		\frac{1}{\taubb}={G_{0\nu}\times \mathcal{M}^2 \times \left(\displaystyle \frac{\mbb}{m_\el}\right)^{\!\!2} }
	\end{equation}
	where the first factor, called the phase space, has the dimensions of an inverse time; the electron mass $m_\el$ 
	is included only for convenience. For a long time, the researchers believed that the uncertainties on $\mathcal{M}$ were as large as a factor of $2-3$ \cite{errolarge}. 
	This view was questioned when the first calculations with error-bars 
	appeared~\cite{faess}, since they exhibited small errors.

	Recently the old view returned in a new form. Indeed, it was shown that the smallness of the error 
	depends essentially upon the assumption that the nucleonic charged current couplings, and in particular the axial 
	coupling $g_A$, are known~\cite{iac}. 
	However, there are good reasons to believe that in the nuclear medium the value for this parameter is not the same 
	as in vacuum, namely $g_A\approx 1.269$, but rather a more `quarkish' value $g_A \approx 1$; 
	moreover, if the analogy with the double 
	beta decay with two neutrino emission holds true, further downward renormalization is expected. 
	The question of the renormalization in the nuclear medium should be considered a priority for this field of research, since ${\mathcal M}$ decreases approximatively as $g_A^2$. As it is clear from  Fig.\ \ref{fig1}, it seems conservative to assume that the uncertainty on the matrix element is not small~\cite{iac}.

	\begin{figure}[t]
		\begin{minipage}[c]{5.7cm}
		\caption{Lifetime for the decay mode of Eq.\ \ref{kmode} that can be probed with a 540\,kton Cherenkov detector and 
			50\,kton Liquid Argon detector, as a function of the data-taking time; the jumps represent the detection of one 
			background event.  
			The arrow indicates the experimental situation in 2007~\cite{cryo}.\label{fig2}}
			\end{minipage}\hskip8mm
			\begin{minipage}[c]{7.0cm}
		\includegraphics[width=6.7cm]{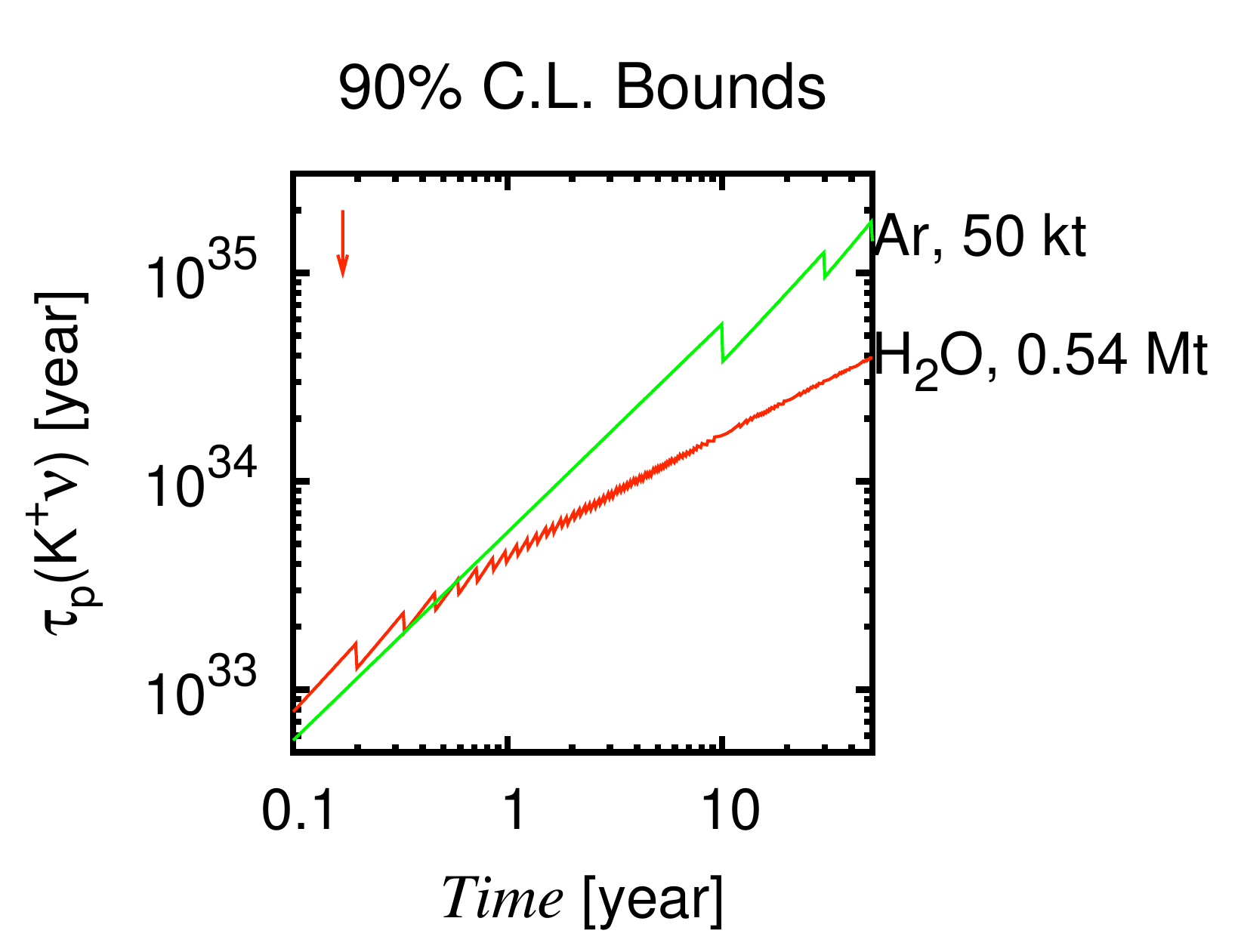}
		\end{minipage}
	\end{figure}

\section{Proton decay?}
	\label{sec:pdecay}

	Here, we complete the discussion about proton decay   and introduce the topic of neutrino physics with large water 
	Cherenkov detectors.

%

\subsection{N.\,D.\,E.\ -- or microphysics with the biggest detectors}

	Various gauge models, such as SU(5) and SO(10), have motivated the search for proton (or nucleon) decay. 
	Very stringent limits have been obtained by KamiokaNDE and Super-KamiokaNDE \cite{sk}; recall 
	that the acronym N.\,D.\,E.\ meant originally 
	Nucleon Decay Experiment. \footnote{Other experiments conceived in the same years, whose names remind us of the prevailing theoretical ideas, are  the NUcleon Stability EXperiment (NUSEX) \cite{nusex}
	and
	the Monopole, Astrophysics and Cosmic Ray Observatory (MACRO)
	\cite{macro}.} 

	After more than 30 years, the predictions remain rather vague, not only due to the nuclear matrix elements, but also due to the 
	incomplete formulations of gauge models. In order to understand the nature of the theoretical difficulties, it is sufficient to think to the Higgs sector of the extended models or to the 
	mass spectra 
	of the various unobserved particles. Notice that a lot of discussion concerned the search for the mode 
	\begin{equation}
		\pr \to K^+ \bar\nu \label{kmode}
	\end{equation}
	motivated by the assumption that supersymmetry is at the TeV scale, where the decay is caused by a loop diagram 
	involving superheavy triplets and supersymmetric particles.
	Of course, the fact that supersymmetry is still unseen does not strengthen the motivations to search for this specific decay mode. 
	
	On the other hand, there are interesting ideas on how to investigate further decay modes as this one with new experiments.
The sensitivity that could be reached 
is shown in Fig.\ \ref{fig2}, from~\cite{cryo}. 
	In water, the efficiency is 14.6\%, obtained summing the 2 methods used by SuperKamiokaNDE,
	and the background rate is $14/(\mbox{Mton}\times \yr)$; 
	in Argon, the efficiency is 97\% and the background rate is $1/(\mbox{Mton}\times \yr)$
	\cite{cryo}.
	Note that the impact of the uncertainty on the background is not included, but at present this is large and it is especially relevant for the Cherenkov detectors, since background events should be soon  observed.
	Summarizing, we see that an Argon detector would permit us to proceed in the zero-background condition for 
	more than an order of magnitude, until the first atmospheric neutrino event will appear (putting aside the possible statistical fluctuations and the discussion of the systematics).

\subsection{Mton water Cherenkov detectors and beyond}

	Independently of the original motivations, Cherenkov detectors reached excellent 
	results by studying atmospheric neutrinos; today, the acronym N.\,D.\,E.\ usually refers to 
	Neutrino Detection Experiment. 
	The HyperKamiokaNDE \cite{hk}, discussed in the contribution of M.\ Shiozawa, will investigate CP by long-baseline 
	oscillations, continuing also the study of atmospheric neutrinos and proton decay, and doing much more: e.g., 
	 monitoring core collapse supernovae.

	The mass of HyperKamiokaNDE will be a fraction of Mton. In this sense, it can be compared with the largest detectors 
	of this type ever conceived/imagined for the study atmospheric neutrinos: the PINGU project in the IceCube 
	site and the possible plan of a similar setup in the Mediterranean sea (ORCA). We will discuss further these detectors in the next section and come back on the IceCube detector in the last one. 

\section{Neutrino oscillations\label{nooco}}

	Oscillations progressed greatly since Pontecorvo's review. The main achievements are
	\begin{itemize}
	\item The theoretical finding  \cite{me}
	that the propagation inside the  matter (e.g., Earth, sun, supernova) modifies  the oscillations. \footnote{In essence, this is an additional refraction effect acting on electron (anti)neutrinos, that works together with the refraction acting on the neutrino mass eigenstates, due to their masses.}  
	This is called ``matter effect''.
	\item The discovery of several anomalies and the 
	confirmation of two of them, leading to the determination of two differences of masses, as mentioned in
	\textsection~\ref{merlos}.
	\item The measurement of the three mixing angles of the ordinary three neutrinos.
	\end{itemize}
	 Here, we discuss the general attitude of this field of research and then focus on the issue of measuring neutrino mass hierarchy. Finally, we briefly discuss the possibility that there are 
	other neutrinos in addition to the three known neutrinos.

	\begin{figure}[t]
		\centering
		\includegraphics[width=9cm]{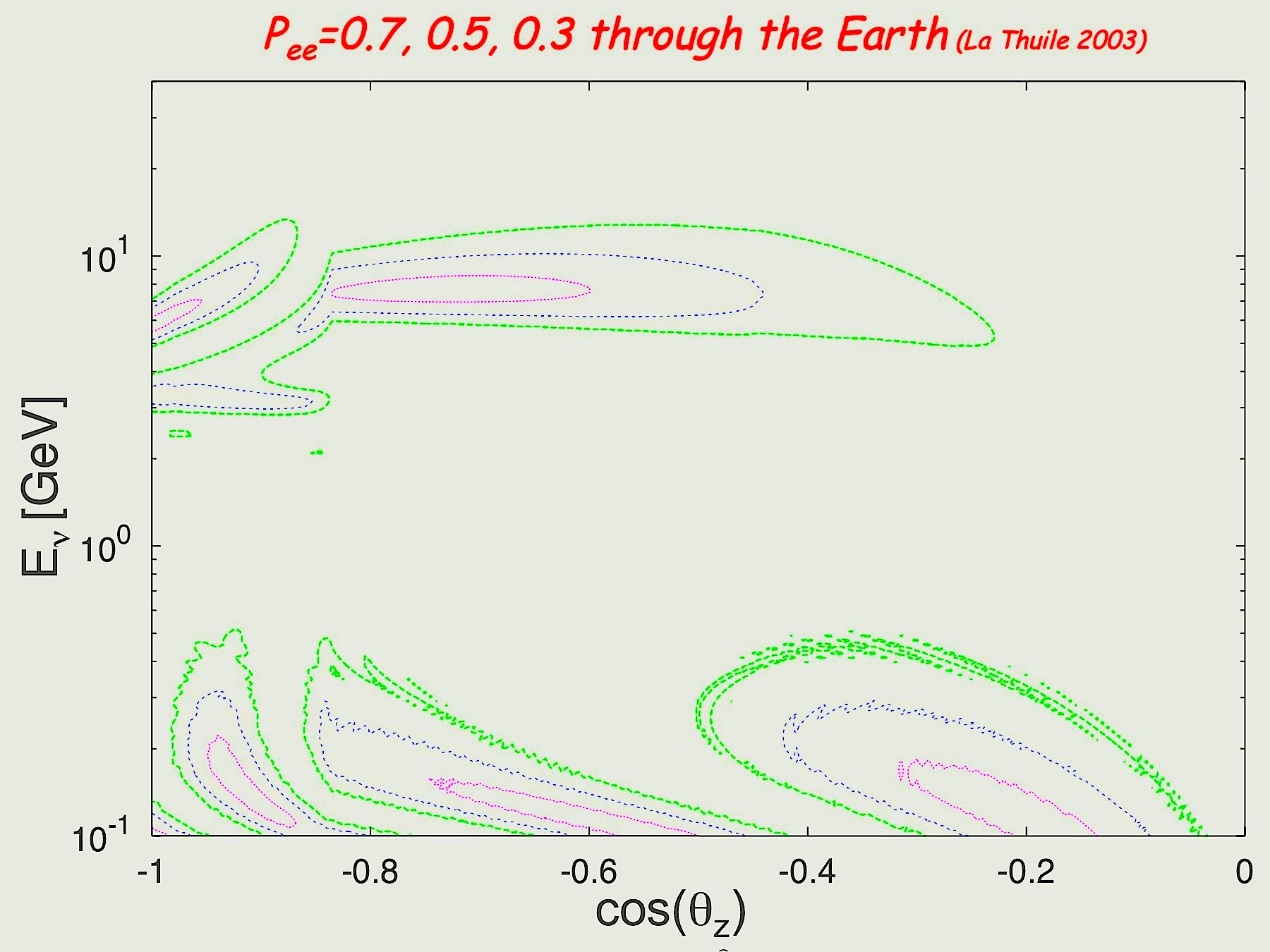}
		\caption{Numerical calculation of $\nu_e$ survival probability $P_{ee}$, for $\nu_e$ produced at the heigth of 15\,km in 
			the atmosphere and crossing the Earth. The regions at about $5-10$ GeV (`atmospheric islands') are those relevant for 
			the study of the neutrino mass hierarchy. 
			The mantle-core discontinuity causes the change at $\cos\theta_Z\sim 0.8$. The 3 `solar islands' above 0.1\,GeV 
			result from an interplay between matter effect and vacuum oscillations; the fuzzy contours are due to 
			$\theta_{13}$ driven vacuum oscillations. From  \cite{lt2003}.}
		\label{fig3}
	\end{figure}

\subsection{Measurements or discovery?}

	The question on whether the study of neutrino oscillations is mainly a way to discover something new or, rather, 
	it is a tool to measure the oscillations parameters of the ordinary neutrinos has a certain importance. 
	In fact, these two attitudes lead to put emphasis on different aspects of the discussion, and consequently, to invest 
	the efforts on different types of experiments. I feel inclined toward the second point of view, as I have described in La Thuile 2003~\cite{lt2003}, but one should be aware that this is not a 
	universal attitude.

	For instance, the idea that $\theta_{13}$ was zero, until the contrary was proved, has been lingering in our field, 
	influencing the discussions. Even now various colleagues like to dub the observational fact 
	$\theta_{13}\sim 9^\circ$ as ``unexpected'' in talks 
	or in written works, often without feeling the need to offer a reference in support of this opinion  
	or to justify such a view with a scientific argument.
	Of course, I would like to contrast this prejudice  with the attitude that $\theta_{13}$ could be just below the 
	experimental limit and possibly within reach, as it was. \footnote{
	Curiously,  this alternative attitude was supported by some theoretical models,
	where this angle was argued to be close to the Cabibbo angle $\theta_C\sim 13^\circ$, 
	see e.g.\ \cite{ecchec}; evidently, other ideas/models/views/prejudices have guided  the discussions.}

	This situation shows that theoretical opinions offer opportunities, but they can also expose us to risks. 
	A healthy research field has to be able to consider a multitude of opinions, rather than passively suffering the 
	effect of the prevailing one \cite{ncim}. Furthermore, it is essential to formulate clearly the motivations of 
	any specific theoretical proposal, at least, just to learn something from its failure.

\begin{figure}[t]
		\centering
		\includegraphics[width=8.6cm]{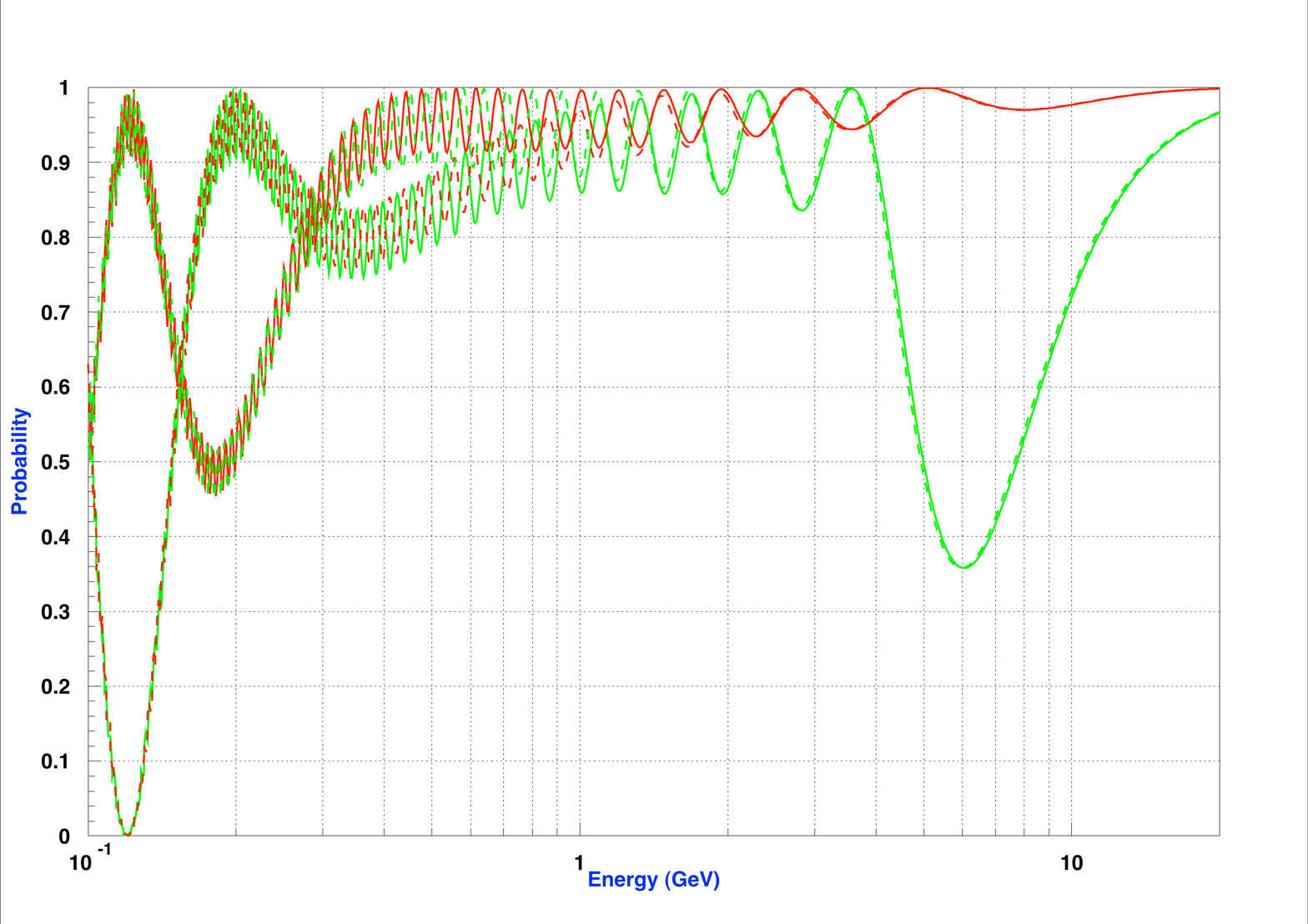}
		\caption{Survival probabilities of $\nu_e$ (green continuous/red dashed) and $\bar{\nu}_e$  
			(red continuous/green dashed) propagating through one Earth radius. 
			Continuous/dashed lines are for normal/inverted hierarchy. 
			The  parameters $\theta_{23}=42^\circ$,  $\theta_{12}=34^\circ$,  $\theta_{13}=9^\circ$,
			$\delta=270^\circ$, $\Delta  m^2_{12}=7.5\times 10^{-5}$ eV$^2$, 
			$\Delta  m^2_{23}=2.4\times 10^{-3}$ eV$^2$. 
			The amplification of $\theta_{13}$ occurs for $\nu_e$ in normal hierarchy/for $\bar\nu_e$ in inverted hierarchy, 
			and it is maximal at  $\sim 7$ GeV. The  wiggles due to $\theta_{13}$ driven oscillations  at lower energies, as the oscillations with solar parameters at even lower energies are also  
			visible. 
			Plot obtained by the web utility \cite{webs}; the agreement with Fig.~\ref{fig3} at $\cos\theta_Z=1/2$ is reasonable.}
		\label{fig4}
	\end{figure}

	It is important to progress toward the measurements of the leptonic CP violating phase and the discrimination of 
	the type of mass hierarchy. 
	I do not believe that the best approach is \emph{to prove that the leptonic CP violating phase is non-zero}, but rather,
	that we should measure this parameter.  I consider remarkable that the first hints \cite{capozzi}, and the existing 
	experimental programs partly discussed in this round table, suggest that this measurement will be successfully achieved 
	in the next years. 
	Comparably, the experimental investigations of the neutrino mass hierarchy by means of oscillations seem to be more demanding. We discuss some of them in the next paragraph.

\subsection{Probing the hierarchy by observing the Earth matter effect}

	One of the reasons to study atmospheric neutrinos is to disentangle the neutrino mass hierarchy question 
	by using the amplification of neutrino oscillations due to   
	 Earth matter. 
	As it can be seen from Fig.\ \ref{fig3} (from La Thuile 2003 \cite{lt2003}) and Fig.\ \ref{fig4}, reasonable conditions for this purpose could be obtained 
	for certain directions and for neutrino energies of many GeV. 
	Indeed it has been argued in \cite{akh} that the discrimination of the mass hierarchy could be possible using 
	PINGU/ORCA like detector. Subsequent more realistic studies pointed out that the required exposure are quite 
	large, of the order of tens of Mton$\times$year, which could however be diminished by distinguishing  
	 track- from shower-events \cite{sense}.

	Another option consists in devising an  
	experiment to maximize the effect. 
	This was implemented in~\cite{epjc} with the use of
		1) a conventional muon neutrino beam from pion decay; 
		2) a muon detector to identify $20-40$ m tracks
		(5\,m in water are $\sim1$\,GeV);
		3)~a pair (source, detector) whose distance maximizes the effect.
	A beam of $6-8$\,GeV sent 
	at ($6,000-8,000$)\,km 
	(e.g., from  Fermilab to Mediterranean 
	Sea or from CERN to Lake Baikal) 
	implies a difference between the two hierarchies of 30\%.
	With $10^{20}$ protons on target and 1\,Mton detector, we have of $\sim1000$ muons. The measurement of 30\% difference is easy and the interpretation unmistakable; furthermore, one can even consider a run with $\bar\nu_\mu$ beam to revert the action of the matter effect.

\subsection{{\em Other} neutrinos}

	After the measurement of the $Z^0$ width \cite{leps}, the proliferation of neutrino species apparently ended. The existence of 	3 invisible fermionic decay channels (i.e., neutrinos) suggests the reliability of a correspondence between 
	leptons and hadrons, proposed long ago.
	However, theorists have hypothesized the existence of
	neutral fermions of all sort of masses and without interactions with the $Z^0$,  
	that could act as neutrinos in various situations. For instance, while cosmic (gravitational) 
	probes agree with the number of 
	standard neutrinos $N_\nu=3$ for what concerns direct cosmological probes such as Planck or primordial 
	nucleosynthesis, other observables, in particular galaxy counts, suggest a somewhat larger number of relativistic 
	species. 

	\begin{figure}[t]
\begin{minipage}[c]{6.9cm}
\caption{The region to interpret LSND in terms of oscillations (yellow), 
		compared with the bounds from other oscillation experiments (dotted line), with the bound from 
		big-bang nucleosynthesis (BBN, red line), with the limit from the neutrino 
		contribution to the energy density (dot-dashed line); recall that 		$\Omega_\nu h^2=\sum_i m_i/93.5\mbox{ eV}$.
		From \cite{cirels}.\label{fig48}}
		\end{minipage}\hskip10mm
\begin{minipage}[c]{7cm}
		\includegraphics[width=5.3cm]{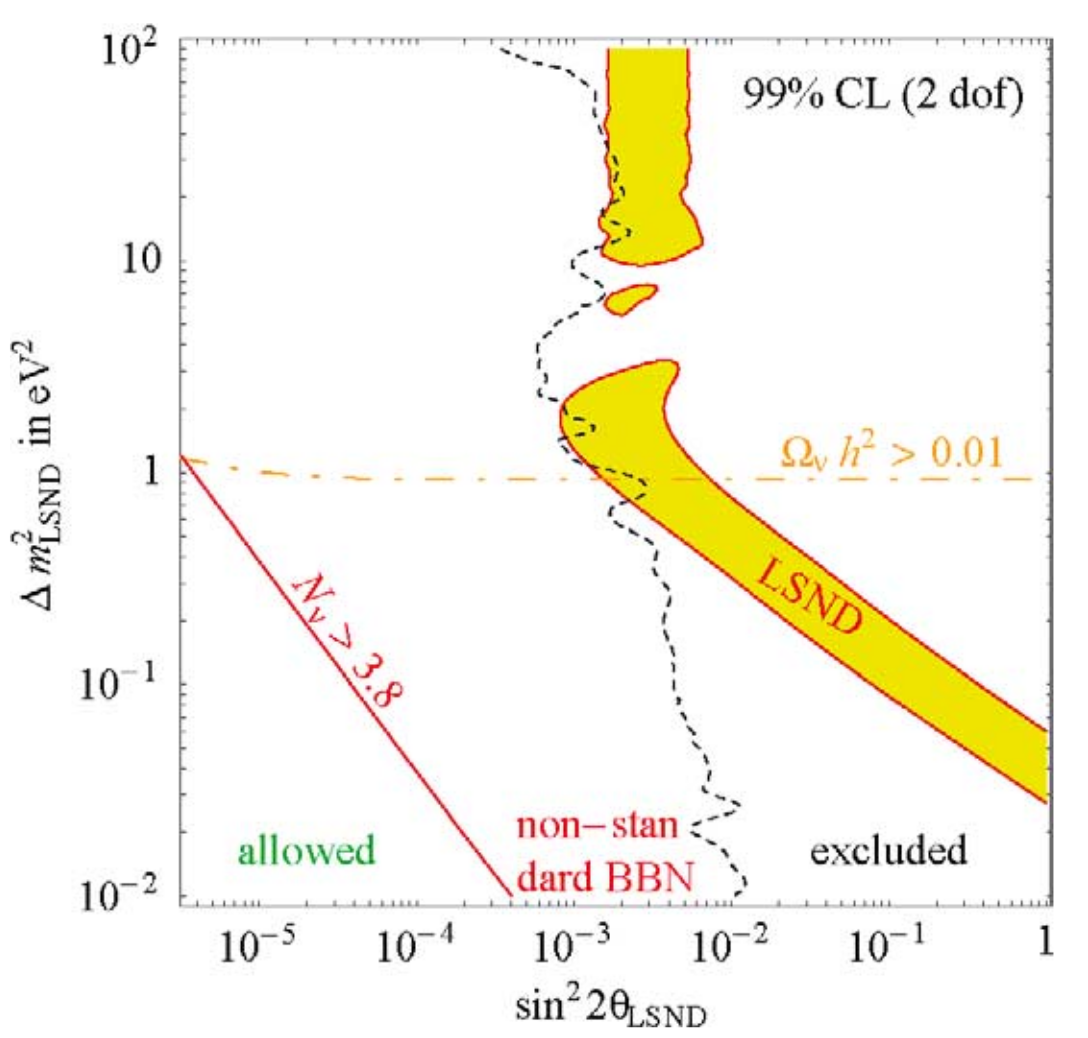}
		\end{minipage}
		\end{figure}

	Even more interesting are the manifestations of these hypothetical neutrinos in oscillation phenomena. 
	In particular, the LSND anomaly~\cite{lsnd} 
	cannot be explained within an oscillation scheme with 3 neutrinos, whose masses are already fixed by other evidences 
	of oscillations. Thus, the existence of additional neutrinos, with masses in the eV range and small mixings with 
	the ordinary neutrinos, has been hypothesized since long. This is widely discussed, also in review papers. E.g.,
	\cite{cirels} outlines the tension of the $N_\nu=3+1$ neutrino interpretation, indicates relevant observables and attempts to 
	identify the critical issues. Instead, \cite{lsn} emphasizes the anomalies coming from reactor, Gallium, cosmology and 
	\emph{``provides motivations for a new round of measurements''}. \footnote{Let us compare  these 
	two works using bibliometric criteria. The former one appeared 10 year ago, it is published and it has 4 authors, 
	49 pages and 11 citations/year. The latter appeared 2 years ago, it is a `white paper' and it has 187 authors, 
	269 pages and 96 citations/year.}    
	Anyway, these anomalies are few $\sigma$ only and a coherent picture does not seem to emerge yet. 
	Things are still evolving and new relevant information should appear soon. In similar cases, it is not easy to plan a new experiment. However, it remains crucial to formulate the scientific questions, that we want to address, as clearly as possible.


\section{Solar, supernova, high-energy cosmic neutrinos}

	Although it is useful to proceed with a separate discussion for \emph{low energy} and \emph{high energy} neutrinos, 
	a very precise definition of the boundary is not necessary for our purposes, and we can set it at 1\,GeV. 
	Well below, there are neutrinos from a large variety of nuclear phenomena; well above, high energy neutrinos 
	plausibly coming from the astrophysical sources of the cosmic rays. In the middle, we have neutrinos from the 
	Earth atmosphere, mentioned above in connection with oscillations. In view of the present interests and discussion, 
	we will in fact focus the discussion on the  neutrinos with the 
	lowest ($\lesssim $ MeV) and highest ($\gtrsim$ PeV) energies.

\subsection{Lowest energy neutrinos / Borexino and beyond}

\begin{figure}[t]
		\centering
		\includegraphics[width=10cm]{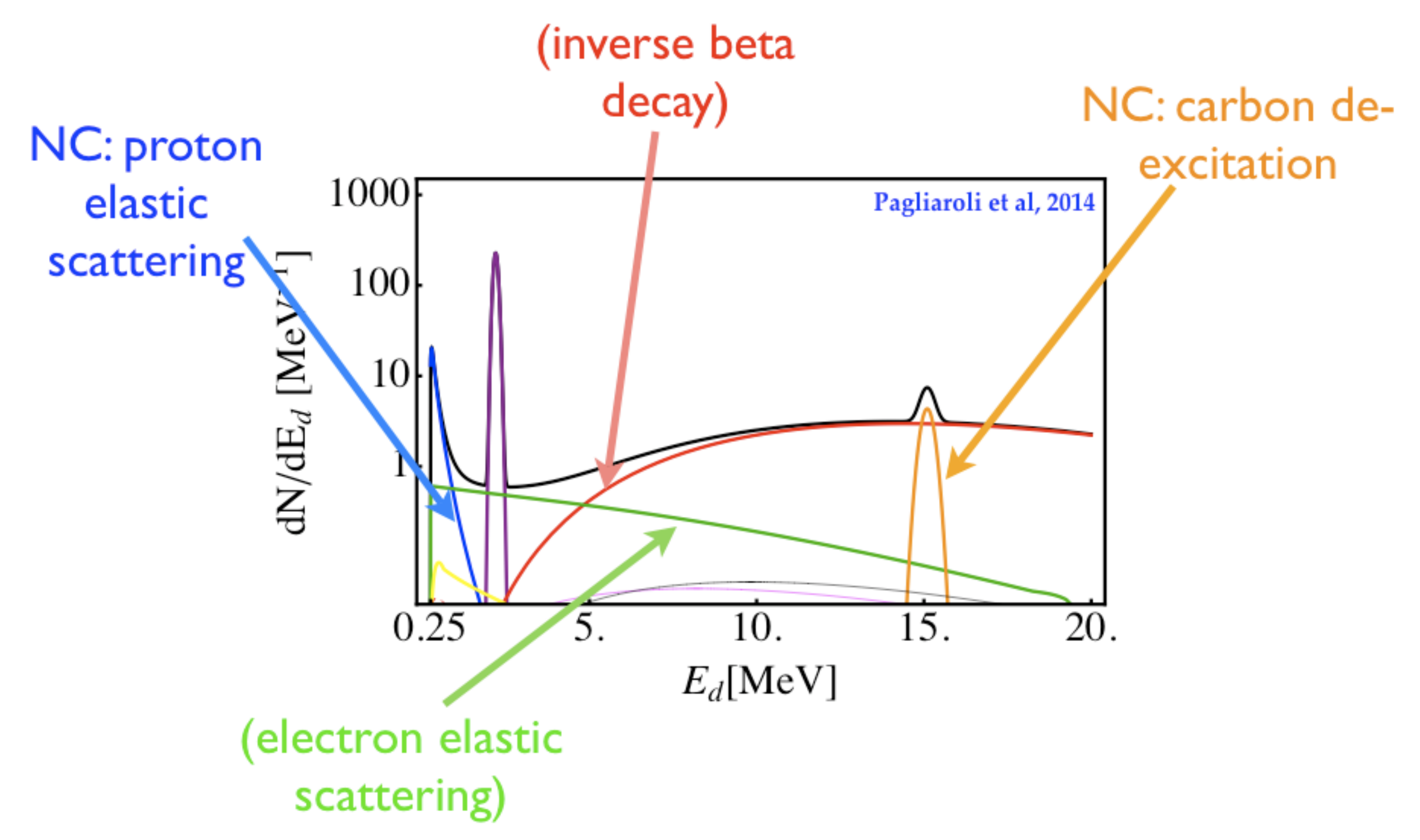}
		\caption{Expected counting rate in Borexino from a galactic supernova at 10\,kpc~\cite{pag}. 
			The two main neutral current (NC) channels are emphasized.}
		\label{fig5}
	\end{figure}

	The success of the Borexino experiment, that achieved unprecedentedly low energy thresholds,  
	has motivated serious consideration about how to proceed with ultra-pure scintillating detectors. 
	Along with Borexino, with KamLAND and with the fore-coming SNO+, this issue concerns the physics with
	much larger detectors, such as LENA and JUNO, of several ten of kton mass \cite{lj}.
	Indeed, there are various scientific reasons why such an extension is interesting. These include:\\ \indent
{\sc Geoneutrinos.} Indeed, up to now Borexino and KamLAND have obtained a relatively small statistics; thus, they have only a moderate power to discriminate the various nuclear chains contributing to the signal~\cite{geoandgeo}.\\ \indent
		{\sc Low energy solar neutrinos.}  We would like to see  in real time the pp neutrinos, 
			but also of the secondary chain of energy production, i.e., those from the CNO cycle. This goal could be achieved in the next 
			years with (a possibly upgraded version of) Borexino. \\ \indent
		{\sc Oscillation studies with reactor neutrinos,} as suggested in~\cite{petcov}. At present, this is the main 
			motivation of JUNO, as discussed in the contribution of J.~Cao.\\ \indent
	{\sc Neutral current events from a galactic supernova.} This category includes the $\nu+\pr \to \nu +\pr$ 
			events that, due to the kinematic of the reaction and to the quenching of the light, select neutrinos of 
			relatively high energies.\\
	Fig.\ \ref{fig5} illustrates the last point~\cite{pag}. 
	Note that a large scintillator detector can contribute to many more physics issues. E.g.,  
	Fig.\ \ref{fig5} shows that the neutral current events can be measured also by the 15.1\,MeV gamma from 
	carbon de-excitation.

	In passing, we offer a remark on a potentially interesting measurement,  namely the detection of relic neutrinos from past supernovae
	(aka, diffuse neutrino background). 
	This requires a good performance in the 
	region of ($20-40$)\,MeV.
	The interaction rate and its uncertainty range  was assessed in~\cite{aea} using 
	the observations of SN1987A, assumed to be a standard supernova, and the astronomical data. 
	The prediction is ($0.01-0.04$) events per kton per year in a water Cherenkov detector, and similarly in a scintillator, 
	which implies that the measurement will be quite demanding.

\subsection{Highest energy neutrinos / the beginning of the IceCube era}
	
	The IceCube experiment has observed events that can be identified as showering neutrinos (i.e., not due to muon tracks) 
	with energies as high as few PeV. This is quite impressive since the center of mass 
	energy of the collision $s\sim 2 E_\nu m_N\sim (2\,\TeV)^2$ is larger than the one directly probed, though it 
	is believed that the partons, at this energy, are known from proton-proton collisions. 
	Moreover, it is quite curious that at these energies the cross section is large enough, that the Earth is not anymore transparent to neutrinos.
	  
	The most recent data have been reviewed by J.\ Auffenberg at this meeting: they are 36 events, of which only 9 are 
	up-going, 8 are tracks (i.e., muon events) one probably spurious. It is quite unlikely that these events are due to atmospheric neutrinos, but in an extreme interpretation of this type, one should assume an unexpectedly large contribution from charm decay (prompt neutrinos). However,  in this case we should   
	not expect any significant flux of tau neutrino. Therefore, an observation of tau events would provide a strong support to the cosmic 
	neutrino hypothesis. 
	
	An alternative approach is offered by the study of the fraction of 
	tracks $f$: In fact, the atmospheric hypothesis would lead to not less that $f=1/2$ muon neutrinos, 
	whereas the cosmic one would lead to $f=1/3$ of muon neutrinos. Suppose to observe $n$ tracks out of $N$ events.  If $f$ is the true (but unknown) fraction of muons,  using
	$\Delta f^2=-\mathcal{L} / \frac{d^2\!\mathcal{L}}{df^2}$ 
	and the likelihood $\mathcal{L}\propto f^n (1-f)^{N-n}$  
	we find
	\begin{equation}
		f=f_*\pm \sqrt{\frac{f_* (1-f_*)}{N}} \quad \mbox{where} \quad f_*=\frac{n}{N}
	\end{equation}
	that in our case reads $f=0.20\pm 0.07 $.
	In order to reach a firm conclusion, a moderate increase in the statistics and a good understanding of the efficiencies 
	of detection will be necessary~\cite{osciPeV}. Note that at the highest energies, it is not easy to measure the muon 
	neutrino energy with IceCube, since the track is too long to be contained in the detector.

	The flux above 60\,TeV is about $E^2 \Phi \sim 10^{-8}\,\GeV/(\cm^2\,\sss\,\sr)$. This is 1/3 as originally estimated 
	by Waxman and Bahcall \cite{wb}, assuming that the energy in cosmic neutrinos equates the one evaluated from the 
	observed highest energy cosmic ray spectrum.
	The flux could be attributed to a variety of cosmic sources, e.g.\ active galactic nuclei or perhaps gamma ray burst 
	(though the search of events correlated in time with the gamma ray burst gave a null result).
	There could be a significant component, at present up to 25\%, due to the region around the center of the Milky Way, even if on statistical 
	basis the events are compatible with a uniform distribution in the sky. 
	Note that if the knee of the all-particle spectrum is a property of the accelerator(s), and if we consider the corresponding neutrino flux, we would expect a 
	cut at 3\,PeV/20 $\sim$ 150\,TeV in the neutrino energy. This is one order of magnitude below the highest energy observed by IceCube.

	\begin{figure}[t]
		\centerline{\includegraphics[width=4.3cm]{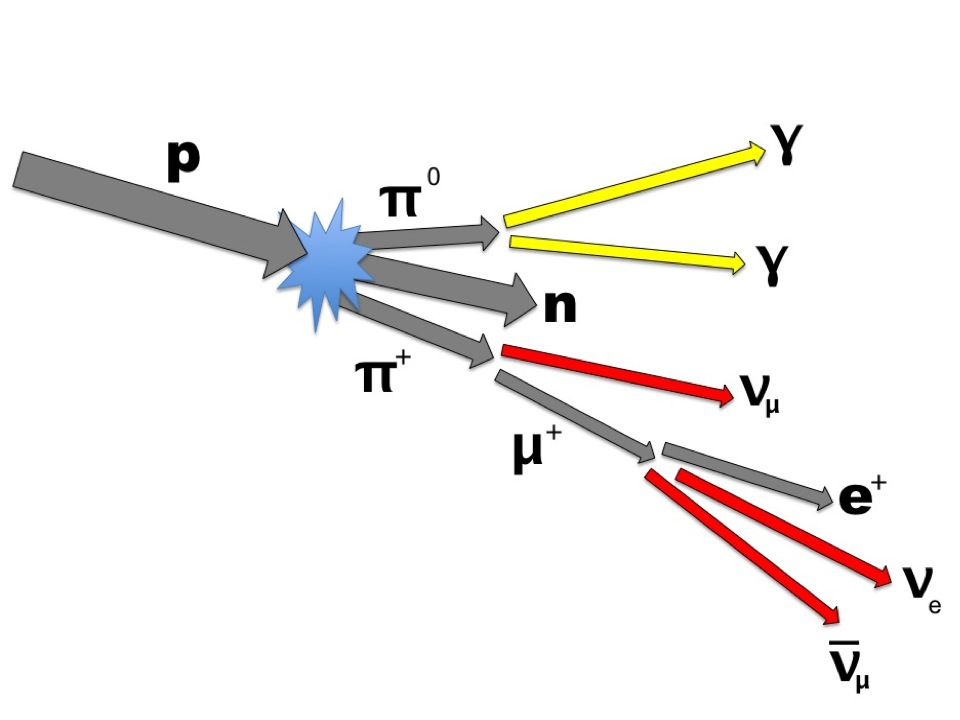}\hskip15mm
		\includegraphics[width=7.7cm]{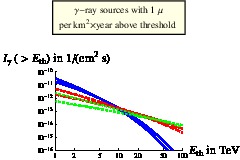}}
		\caption{Left: illustration of the expected link between gamma rays and neutrinos, in the assumptions that these are 
			both generated by cosmic ray collisions and that the gamma rays are unchanged after production. 
			Right: the inclusive flux of gamma that, in the same hypotheses, leads to a flux of a muon per km$^2$ per year 
			above 1 TeV. From \cite{flx}.}
		\label{figpp}
	\end{figure}

	These exciting results also renew the interest  in other goals of high energy neutrino astronomy. In particular, 
	it would be important to know the neutrino flux from galactic sources, either from a relatively wide object as 
	the galactic center, already mentioned, or from quasi-point sources, as supernova remnants, molecular clouds illuminated 
	by cosmic rays, pulsar wind nebulae, etc.\,. 
	As discussed in this meeting, this motivates adding an extended cosmic ray veto (surface detector) in IceCube and it is  of great interest for neutrino telescopes located in the 
	Northern hemisphere, as Antares and the future Km3NET, both in the Mediterranean Sea, or NT1000 in Lake Baikal. 
	
	The search of galactic high energy neutrinos is tightly linked, in many cases, with the search of high energy gamma rays, as illustrated in Fig.\ \ref{figpp}.
	Indeed, it has been shown that all gamma ray sources transparent to their gamma rays should release at least 
	$(1-2)\times 10^{-13}$ photons per $\cm^2$ per second above 20\,TeV, in order to produce at least 1 muon per km$^2$ per 
	year  \cite{flx}. This remark points to an important synergy with the future gamma ray experiments that aim to measure in this window of energy, as LHAASO, or the large set of small telescopes of CTA.

\section{Conclusive remarks}

	 Rather than attempting a provisional summary, that is probably of little or of no interest, 	I would like to conclude by returning on 
	 general considerations.
	
We could say that the review  of Pontecorvo \cite{bruno}
concludes the pioneering stage of neutrino physics; but the progresses obtained in the subsequent thirty years, that we recalled only partly, are just impressive. 
We live in an exciting moment, however in order to plan at best the next steps forward, it is important to remain aware that the progresses are not only due to new and lucky circumstances, but also to certain specific features of this field of research. 
Among these positive features, we have the lively connection between theory and experiment.

It is quite evident that neutrino physics is assuming more and more the features of a `Big Science', that implies large social  groups and their characteristics dynamics: specialization, hierarchical organization, consensus,  etc.\,.
These dynamics can have major effects on the field:  e.g., 
increasing the separation between theory and experiment, 
encouraging an excess of speculative attitude among  theorists, \footnote{I love 
the appeal to realism made by a character of the TV series Futurama
in a dialog of `Mars University' episode: 
[Professor Farnsworth:] {\em Nothing is impossible if you can imagine it! That’s what being a scientist is all about!} [Cubert:] {\em No, that’s what being a magical elf is all about.} 
It is amusing that, in the same episode,  neutrinos are mentioned.}
favoring the narrowing of the research fields or worse the lack of 
innovation. 
(Apparently, Pontecorvo himself had similar worries, as it is clear from his words \cite{bruno}: {\em The expenditure of resources has been justified, but one should neither underestimate the importance of high-energy neutrino physics, nor overestimate it. This is not pessimism, but an appeal to avoid routine.})

I consider of vital importance succeeding to maintain an active link between theory and experiment. This is needed not only to raise new questions and working hypotheses, but also to provide occasions of confrontation, of doubts and also of   contradictions.  Indeed, science requires a continuous assessment of the validity of the assumptions that are adopted and of the conclusions that are reached. This is possible if the scientific community is informed, competent and concerned, but also open-minded,  diversified, and better if continuosly renewed.


In my humble opinion, we should succeed in recognizing the increased role of various  practical (political, economical)  considerations in neutrino physics, but without relegating the scientific debates to a marginal role. It is  advisable that the decisions that concern the future of  this field  remain based on open and frank discussions among scientists, and I am glad to the Organizers for the occasion offered by this round table.

\acknowledgments

	I thank G.~Bellettini, G.~Chiarelli, M.~Greco, G.~Isidori for invitation and support, together with  
	{\rm M.\ Chizhov, S.\ Dell'Oro, E.~Lisi, S.\ Marcocci, G.~Pagliaroli, S.\ Recchia, F.~Terranova} 
	for useful discussions. collaboration and help.


\section*{Questions/remarks}

	\noindent
	{\sc Studenikin:} The study of magnetic moments is potentially important to investigate the nature of neutrinos.
	\\
	{\sc FV:} This is certainly true, even if I am not aware of a convincing theoretical case or a clear observational 
	hint for large magnetic moments. I suspect that supernova neutrinos could be considered valuable in this respect, 
	despite the difficulties in modeling the relevant astrophysics, owing to the very intense magnetic fields. 
	\\[2ex]
	\noindent
	{\sc Bellettini:} How large is the Earth matter effect and what it is its interplay with the other effects?
	\\
	{\sc FV:} 
	This depends critically upon the specific experimental conditions. See e.g., the talk of Shiozawa for the case of 
	Hyper-Kamiokande. The matter effect has been maximized and 
	other effects, in particular leptonic  CP violation,  do not play a significant role 
	for the specific setup I considered above (i.e., muon detector + pion beam)~\cite{epjc}.
	\\[2ex]
	\noindent
	{\sc Auffenberg:} Don't you think that cosmic  neutrinos (in particular the PeV ones, that we observe in IceCube) 
	belong to astrophysics as much as to particle physics?
	\\
	{\sc FV:}  Absolutely yes, and I am not sure that we have much to gain in separating artificially one aspect from 
	the other. 
	In particular, I am not convinced that we can learn much on neutrino oscillations, on the contrary I am sure we can 
	learn a lot {\em from} oscillations~\cite{osciPeV}.
	\\[2ex] 
	\noindent
	{\sc De Rujula:} I would like to stress: $(i)$ the importance of the laboratory search for neutrino masses;
	$(ii)$ the interest to see directly the cosmic neutrino background.
	\\
	{\sc FV:} Thanks. I discussed these points a bit in the occasion of my previous talk at La Thuile (2003), 
	however $(i)$ I agree that I did not emphasize sufficiently the first point in this talk, especially in view of the 
	facts that KATRIN is almost to produce data, and that HOLMES offers interesting prospects \cite{hol}; $(ii)$ as for the second point, 
	I believe that the most urgent matter is to make sense of the cosmological data, for what concerns the role of neutrinos 
	and of their masses. The progress of sensitivity is impressive but caution in the interpretation is in my view necessary. \footnote{20 year ago, the hot+cold matter cosmology indicated neutrino masses of 
	$\approx $2.4\,eV \cite{20}; 10 year ago,  
	$\sum_i m_i\approx $0.6\,eV  \cite{10}; today the hints for non-zero neutrino masses are half as small~\cite{00}.}
	\\[2ex]
	\noindent
	 {\sc Agostini:} What do we know from the `black-box theorem' on neutrinoless double beta decay?
	 \\
	{\sc FV:}  Not much: the value of $\mbb$, expected from model independent loops, is of the order of 
	$10^{-24}$\,eV~ \cite{lindner}; 
	moreover, neutrinos could be Majorana particles  even if $\mbb$ is zero and the neutrinoless 
	double beta decay is  due to other mechanisms, see e.g.~\cite{mitra}. 
	\\[2ex]
	\noindent
	 {\sc Bertolucci:} How deep should be a neutrino decay tunnel of a 6,000\,km long-baseline?
	 \\
	{\sc FV:}  The required neutrino energy is 3 times smaller than the one of the  CNGS beam, thus a tunnel of half a km should suffice. 
	For an inclination of 30 degrees, this means 250 m depth. (LNGS engineers told me that tunnel boring machines--TBM--can excavate such an inclined tunnel, its cost being about 10 Meuro.)

\end{document}